\documentclass{PoS}

\title{An overview of the EXTraS project: Exploring the X-ray Transient and Variable Sky}

\ShortTitle{The EXTraS project}

\author{ 
\speaker{A. De Luca}$^{a}$\thanks{Primary coordinator of EXTraS},
R. Salvaterra$^{a}$, 
A. Tiengo$^{b}$, 
D. D'Agostino$^{c}$, 
M.G. Watson$^{d}$, 
F. Haberl$^{e}$ and 
J. Wilms$^{f}$ \emph{on behalf of the EXTraS collaboration}\\
\llap{$^{a}$}INAF-Istituto di Astrofisica Spaziale e Fisica Cosmica di Milano (IASF)\\ 
Via Bassini 15, 20133 Milano, Italy\\
\llap{$^{b}$}Istituto Universitario di Studi Superiori di Pavia (IUSS)\\
Piazza della Vittoria 15, I-27100 Pavia, Italy\\
\llap{$^{c}$}CNR-Istituto di Matematica Applicata e Tecnologie Informatiche (IMATI)\\
Via de Marini 6, I-16149 Genova, Italy\\
\llap{$^{d}$}University of Leicester, Department of Physics and Astronomy\\ 
Leicester, LE1 7RH, UK\\
\llap{$^{e}$}MPG-Max Planck Institut f{\"u}r extraterrestrische Physik\\
Giessenbachstrasse, D-85748 Garching, Germany\\
\llap{$^{f}$}Erlangen Center for Astroparticle Physics (ECAP)\\
Sternwartstrasse 7, D-96049 Bamberg, Germany\\
E-mail: \email{deluca@iasf-milano.inaf.it}, \email{ruben@iasf-milano.inaf.it},
\email{andrea.tiengo@iusspavia.it}, \email{dagostino@ge.imati.cnr.it}, \email{mgw@leicester.ac.uk},
\email{fwh@mpe.mpg.de}, \email{joern.wilms@sternwarte.uni-erlangen.de}}

\abstract{The EXTraS project (``Exploring the X-ray Transient
and variable Sky'') will harvest
the hitherto unexplored temporal domain information buried
in the serendipitous data collected by the European Photon
Imaging Camera (EPIC) instrument onboard the ESA XMM-Newton
X-ray observatory since its launch. This will include a search
for fast transients,
as well as a search and characterization of variability
(both periodic and aperiodic) in hundreds of thousands
of sources spanning more than nine orders of magnitude
in time scale and six orders
of magnitude in flux.  X-ray results will be complemented
by multiwavelength characterization of new discoveries.
Phenomenological classification of variable sources
will also be performed. All our results will be made available to the community.
A didactic program in selected High Schools in Italy,
Germany and the UK will also be implemented. 
The EXTraS project (2014-2016), funded within the EU/FP7
framework, is carried out by a collaboration
including INAF (Italy), IUSS (Italy), CNR/IMATI (Italy),
University of Leicester (UK), MPE (Germany) and ECAP
(Germany).}

\FullConference{Swift: 10 Years of Discovery\\
		 2-5 December 2014\\
		 La Sapienza University, Rome, Italy}

\begin{document}

\section{Variability, serendipity and XMM-Newton/EPIC data.}
\label{sec:1}

Almost all astrophysical objects, from nearby stars to supermassive
black holes at cosmological distances, display a characteristic 
variability in flux and spectral shape on a broad range of time scales.
This is especially true in the high energy range of the electromagnetic
spectrum. Study of such variability can yield crucial clues to the 
nature of the sources and to their emission physics -- from magnetic reconnection
in stellar atmospheres to accretion on compact objects.

In the soft X-ray energy range (0.1-10 keV), focusing telescopes
collect each day a very large amount of information about serendipitous
sources and their time variability -- each detected photon is time-tagged! 
-- but such information remains mostly unused in data archives. 
This is the case for the European Photon Imaging Camera (EPIC) instrument 
onboard the ESA XMM-Newton X-ray observatory. EPIC
consists of a pn \cite{strueder01} and of two
MOS cameras \cite{turner01}. Thanks to its unprecedented
combination of large sensitivity to point sources, large field of view, 
good angular, spectral and temporal resolution, EPIC is the most powerful 
tool to study the variability of
faint X-ray sources. Launched in 1999 and still fully operative, EPIC collected
more than 230 Millions of seconds of data so far (with the prospects for 
several more years of observations). Large efforts are ongoing 
to study the serendipitous content of the EPIC database. 

\begin{itemize}

\item {{\bf The XMM-Newton Serendipitous Source Catalogue.} A catalogue 
of all the X-ray sources detected in EPIC pointed observations is periodically
released by the XMM-Newton Survey Science Centre consortium 
\cite{watson09}. The most recent
release, called 3XMM\footnote{http://xmmssc-www.star.le.ac.uk/Catalogue/xcat\_public\_3XMM-DR4.html}, 
is the richest source catalogue in the soft X-ray range 
ever compiled, including more than 530,000 detections of more than 370,000 
unique sources on 800 square degrees of the sky (more
than 65,000 sources have multiple-epoch detections due to the
overlap of different observations). Light curves with uniform time bins were extracted 
for about 120,000 bright sources and basic variability tests were performed.}

\item{{\bf The XMM Slew Survey.} Data collected while the telescopes move 
from one target to the next scheduled one
-- the so-called slews -- are also a very rich resource. 
Even if exposure time is short (7-10 s per source), slews
are as sensitive as the ROSAT All Sky Survey (RASS) below 2 keV, and have
an unrivalled sensitivity above 2 keV, if compared to any currently
available all-sky survey. 
The XMM Slew Survey (XSS) Catalogue \cite{saxton08,warwick12}, 
generated using EPIC/pn slew data, in its most recent release lists 
more than 20,000 detections\footnote{http://xmm.esac.esa.int/external/xmm\_products/slew\_survey/xmmsl1d\_ug.shtml} 
and covers more than 60\% of the sky, 
about 20\% having been scanned more than one time.   
Although thousands of sources prove to be variable when compared
to the RASS, no systematic study of variability has ever been carried out.}
\end{itemize}

In spite of such large efforts, temporal domain information 
in EPIC serendipitous data remained mostly unexplored so far.
The EXTraS project will fill the gap, as a service for the astrophysical community.

\section{A time-domain treasure hunt in EPIC data.}
\label{sec:2}

{\bf EXTraS} -- {\bf E}xploring the {\bf X}-ray {\bf Tra}nsient and variable {\bf S}ky 
-- aims at fully exploring and disclosing the serendipitous content of the EPIC database
in the temporal domain and to make it available and easy to use to the whole community.
EXTraS starts from (and partially builds upon) the work that lead to the release
of the XMM serendipitous source catalogue and of the XMM Slew Survey catalogue.

\begin{figure}
\includegraphics[width=\textwidth]{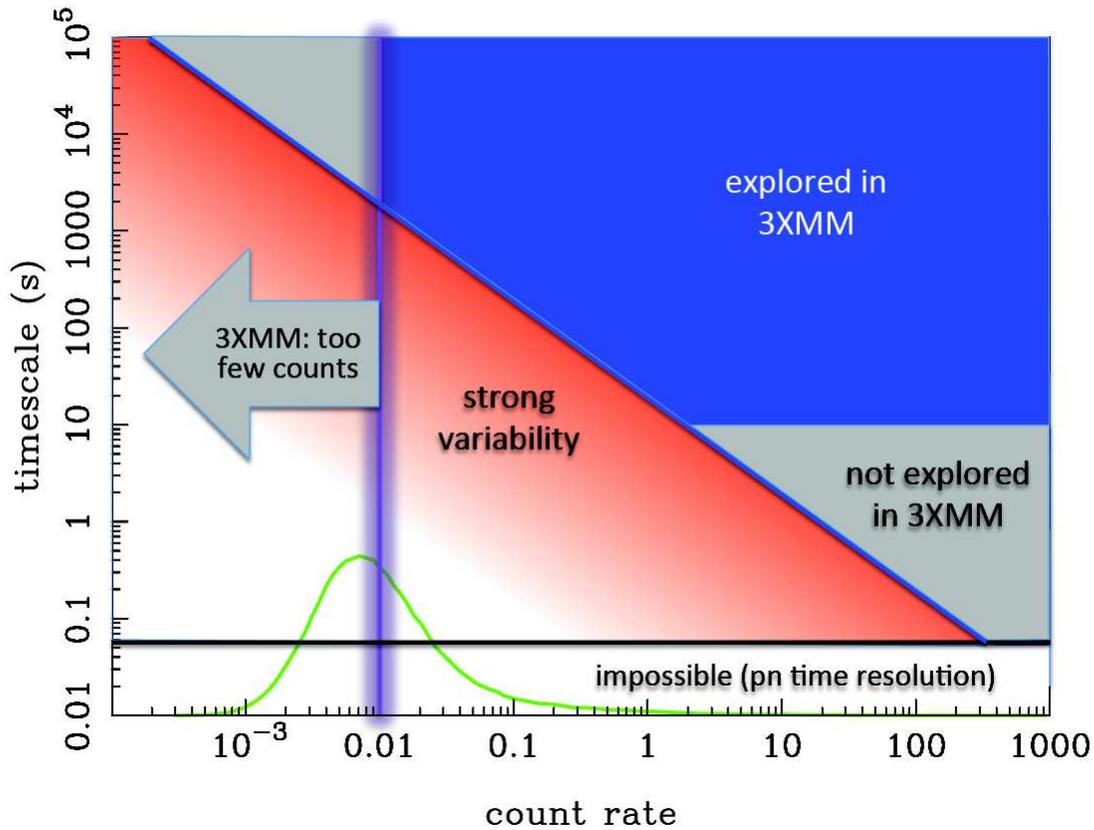}
\caption{Timescale -- count rate parameter space. The blue area marks the region where 3XMM light 
curves with a uniform time-binning have been extracted. The area is limited by the prescription 
used to produce such light curves: (i) the diagonal line to the lower left marks the minimum 
of 20 counts per time bin adopted in 3XMM; (ii) the vertical violet line to the left indicates the 
approximate source count limit for 3XMM light curve creation; (iii) the horizontal line 
marks the minimum time bin of 10 s in 3XMM light curves. The green curve superimposed to the diagram
shows the distribution of EPIC count rates for 3XMM sources. }
\label{fig1}
\end{figure}

Figure~\ref{fig1} shows the time scale vs. count-rate plane and the region 
accessible thanks to 3XMM products. EXTraS will systematically extend such
region. First, we will study the variability of many more sources. 
Second, we will go on much shorter time scales -- by construction, in 3XMM 
a source with a count rate of 0.01 cts/s cannot be studied at temporal 
scales below 2000 s (see Fig.~\ref{fig1}). 
Third, EXTraS will explore for the first time variability on longer time scales, 
beyond the duration of pointed observations. Fourth, we will also search for periodicities. 
Different lines of analysis are being implemented in the project. 

\begin{itemize}

\item{{\bf A systematic study of aperiodic, short-term variability} -- EXTraS aims at characterizing
the bulk of 3XMM sources (possibly about 300,000 sources)
on all possible time scales, from the duration of 
an observation to the instrument time resolution (typically 73 ms and 2.6 s for the pn and MOS cameras). 
The very broad range of count rates and observing conditions 
in 3XMM, together with the wide diversity of aperiodic phenomena to be characterized 
require the use of different techniques to extract all available information. 
Different approaches are being implemented and tested, to generate, e.g., power spectra 
based on Fourier analysis, or optimally-binned light curves based on Bayesian Blocks 
analysis. A whole set of synthetic parameters are then extracted for each source 
in order to quantify and characterize different manifestations of variability. The challenge
is to properly account for several sources of systematics, ranging from gaps in the
time series data, to the highly variable EPIC particle background, to the superimposition
of the signal of different sources.}

\item{{\bf A systematic search for periodicity.} Such a kind of search had 
never been performed before on the EPIC database. Different algorithms will 
be applied to photons' times of arrival, in order to search for pulsations
down to P$\sim0.2$ s in the largest possible number of 3XMM sources
(about 300,000). 
We will use a generalization
of the FFT technique accounting for non-poissonian noise components, as
well as other techniques such as e.g. the Rayleigh test. A characterization
of the signal of all candidate pulsators will be provided. It is also
planned to compute upper limits to the pulsed fraction of all analysed sources.}

\item{{\bf A blind search for transient sources.} This will allow to discover
a potentially large number of sources that rose above the detection threshold for just
a small interval of time. Short transients with a low fluence would easily be missed, 
confused in the background, in any standard
analysis carried out on the time scale of the whole observation. This is the case for the 
analysis used to produce the serendipitous source catalogue -- thus, such transients 
would not appear in 3XMM.
Searches for transients of any duration (possibly down to a few seconds)
will be performed within EXTraS, e.g. using a Bayesian Blocks analysis 
to optimally select promising time intervals for transient source detection. 
A thorough temporal study, down to the instrument
time resolution, will be carried out for each candidate transient, together with a spectral 
characterization. This will also include a search for multiwavelength counterparts, 
mainly based on existing databases.}

\item{{\bf A systematic investigation of long-term variability.} EXTraS will take
advantage of the large number ov overlapping observations performed at 
different epochs. Although at lower sensitivity, information will also
be extracted on the large fraction of the sky visited by multiple slews.
For each 3XMM and XSS source, all count rate measurements at different epochs
will be complemented by upper limits computed in case of non-detections 
in spatially overlapping observations, to produce long-term light curves. 
This will allow to search for -- and characterize -- variability 
on a huge number of sources on time scales up to more than a decade.}

\end{itemize}

Phenomenological classification of all variable sources 
will also be performed by EXTraS, as an important objective of the project. 
We will use an automated statistical classification approach,
based on a full set of source ``features'', describing all X-ray temporal
and spectral properties, as well as on all available information
in multiwavelength catalogues and databases. We will test different 
classification algorithms,
starting from results and lessons learned in ongoing efforts related
to large optical surveys.

\section{The EXTraS output for the community.}
All EXTraS results and products, together with new software tools, 
will be released to the community at the end of the project (end of 2016).
Quality control will be a crucial task within EXTraS. A careful screening 
and validation process will be carried out to reject possibly flawed results
and products generated by our automatic pipelines. A repository will be deployed,
with an easy-to-use, Virtual Observatory-compliant web interface. Highly
customizable interrogations will be possible for all
users, even if they have no previous experience with X-ray data.

The EXTraS database will include a full characterization of the temporal
behaviour of hundreds of thousands
of sources, spanning more than nine orders of magnitude
in time scale (from $<$1 s to $>$10 yr) and six orders
of magnitude in flux (from $\sim10^{-9}$ to $\sim10^{-15}$ erg cm$^{-2}$
s$^{-1}$ in 0.2-12 keV). We trust this will be a very rich and useful
resource for the community, with a direct relevance for a very broad
range of topics in almost all fields of astrophysics -- from strong gravity,
to accretion-related phenomena, 
to magnetic field generation and dynamics, to the solar-stellar connection and the 
habitability of planetary systems. Very different ways of exploitation
can be expected as well, from the study of specific variable sources, 
to population studies of different source classes, to the search
for rare events, to the selection of extreme objects. 
Totally unexpected discoveries can also be foreseen, as has always
been the case when new regions in parameter space have been explored.
A concise overview of possible science cases that could benefit from
EXTraS results is given by \cite{deluca15}.

\section{Education with EXTraS}
We will also implement an experimental didactic program in High Schools
in Italy, Germany and the UK, directly involving students in the 
project activities. After a series of introductory lessons,
students will have access to a selected sample of EXTraS results.
Using our visualisation tools, they will be asked to classify
sources according to their phenomenological affinity to a set of templates.
Results provided by students will be compared to results provided
by astronomers of the EXTraS team as well as to the output of 
our automated classification algorithms.
Thus, our educational activity will turn also to an experiment of citizen science, 
allowing us to assess the viability of involving non-expert (but trained) 
people in a complex classification task.

\section{The EXTraS consortium.}
The EXTraS project is carried out by an international collaboration featuring:

\begin{itemize}

\item{National Institute for Astrophysics (INAF, Italy), coordinating the EXTraS consortium, 
participating mainly with the Istituto di Astrofisica Spaziale e Fisica Cosmica (IASF Milano)
and with the astronomical observatories of Roma (OA-Roma), Brera (OA-Brera), Catania (OA-Catania) 
and Trieste (OA-Trieste).}

\item{Istituto Universitario di Studi Superiori di Pavia (IUSS, Italy);}

\item{National Counseil for Research (CNR, Italy), participating with the Institute of 
Applied Mathematics and Information Technologies (IMATI);}

\item{University of Leicester (United Kingdom);}

\item{Max Planck Society for the Advancement of Science (MPG, Germany), participating
with the Max Planck Institute for Extraterrestrial Physics (MPE);}

\item{Friedrich-Alexander Universitat Erlangen-Nuremberg (Germany), participating 
with the Erlangen Centre for Astroparticle Physics (ECAP).}

\end{itemize}
 
For more information, updates on the status of the project, and a full list of contacts, 
please refer to the EXTraS website, online at {\bf www.extras-fp7.eu}.

\vskip 1cm
EXTraS has received funding from the European Union's Seventh Framework Programme
for research, technological development and demonstration under grant agreement n. 607452.


\begin{thebibliography}{99}

\bibitem{deluca15} De Luca, A., Salvaterra, R., Tiengo, A.,
in proceedings of \emph{The Universe of Digital Sky surveys}, in press
\bibitem{saxton08} Saxton, R.~D., Read, A.~M., Esquej, P., et al.
\emph{The first XMM-Newton slew survey catalogue: XMMSL1}
\emph{A\&A}, {\bf{480}} (2008), 611 (arXiv:0801.3732) 
\bibitem{strueder01} Str{\"u}der, L., Briel, U., Dennerl, K., et al.
\emph{The European Photon Imaging Camera on XMM-Newton: The pn-CCD camera.}
\emph{A\&A}, {\bf{365}} (2001), L18
\bibitem{turner01} Turner, M.~J.~L., Abbey, A., Arnaud, M., et al.
\emph{The European Photon Imaging Camera on XMM-Newton: the MOS cameras.}
\emph{A\&A}, {\bf{365}} (2001), L27 (arXiv:astro-ph/0011498)
\bibitem{warwick12} Warwick, R.~S., Saxton, R.~D., \& Read, A.~M.
\emph{The XMM-Newton slew survey in the 2-10 keV band.}
\emph{A\&A}, {\bf{548}} (2012), 99 (arXiv:1210.3992) 
\bibitem{watson09} Watson, M.~G., Schr{\"o}der, A.~C., Fyfe, D., et al.
\emph{The XMM-Newton serendipitous survey. V. The Second XMM-Newton serendipitous source catalogue}
\emph{A\&A}, {\bf{493}} (2009), 339 (arXiv:0807.1067)
 
\end{thebibliography}
\end{document}